\documentclass{PoS}

\title{Comparison of different non-linear equations \\
             and corresponding unitarization schemes}

\ShortTitle{Comparison of different non-linear equations ...}

\author{\speaker{O.V. Selyugin} \\ %\thanks{}\\
          Bogoliubov
 Laboratory of Theoretical Physics, JINR, 141980, Dubna, Russia \\
        E-mail: \email{selugin@theor.jinr.ru}}

\author{J.-R. Cudell\\
         Institut de Physique, B\^at. B5a,
Universit\'e de Li\`ege, Sart Tilman, B4000
  Li\`ege, Belgium\\
        E-mail: \email{JR.Cudell@ulg.ac.be}}

\abstract{Different forms of non-linear equations
   which mimic parton saturation in the non-perturbative regime
   are examined. These equations lead to  corresponding unitarization
   schemes in the impact parameter representation of the hadron scattering
   amplitude. It is shown how specific properties of the non-linear
  equations reflect  different features of the diffraction processes. }

\FullConference{DIFFRACTION 2006 - International Workshop on Diffraction in High-Energy Physics \\
         September 5-10 2006\\
         Adamantas, Milos island, Greece}

\begin{document}

%\section{...}

\section{Introduction}

   Future experiments at the LHC will open a new page in the physics 
of hadron-hadron
interactions. The new energy region promises  many new physical phenomena.
It is very likely that in this region the asymptotic properties of 
the hadron-hadron interaction will appear.   These properties are connected
with the first principles of the theory such as analyticity, unitarity 
and Lorenz invariance.
In the momentum transfer representation,  some asymptotic bounds were
  obtained many years ago. For example, the AKM (Auberson-Kinoshita-Martin) \cite{AKM} theorem
  leads to possible oscillation at very small $t$ in the elastic differential cross sections.
Such oscillations may have already been discovered in the UA4/2 experiment \cite{ua42,sel-ua42}.
   In the impact parameter representation, a more complicated picture appears. For example,
   the Froissart-Martin theorem based on the first principles establishes
   a bound on the growth of total cross sections
    \cite{frois} - they cannot grow faster than $\log^2(s/s_0)$. However, the Regge
  behavior and the linear BFKL equation lead to a power growth of the total cross sections.
  Obviously, this conflict requires a non-linear regime at super-high energies.
  This regime may be connected with the
  saturation of the density of the partons in
  hadron  and can be related to the saturation of the unitarity bound only
  in the impact parameter
  representation.

 The usual non-linear equations, such as the BK equation \cite{BK},
 are connected with perturbative and  non-perturbative infrared effects that
 lead to the saturation regime at large impact parameters. Hence, they are not
 directly related to the Black Disk Limit (BDL)-saturation which in most models
  is connected  with small impact parameters.
  Also, the models
 based on the color dipole representation of the high-energy interaction \cite{muller}
  suppress, for the most part, the soft hadron interaction. This is the reason why these models
  describes the hard interactions well. On the other hand, the BDL suppresses in most cases the hard
  component. It leads to an increase in the role of the soft component at super-high energies.

 Unitarity of the scattering matrix,
$SS^{+} \ = \ 1$,
is connected with
the properties of the scattering amplitude in the impact-parameter
 representation as the latter is equivalent at high energy
to a decomposition into partial-wave amplitudes.
   At high energies the scattering amplitude in the impact-parameter
 representation can saturate the unitarity bound of the overlapping function
 $G(s,b)$ below some impact parameter
 $b_i$.
One can impose the unitarity condition through various schemes.
Two of them are based on an analytic implementation of unitarity:
  the first is the standard eikonal representation
\begin{eqnarray}
  T(s,t) \ \sim \ i \int_{0}^{\infty}  \ b\  db \ J_{0}(b q) \
     [1 \ -  \  \exp(-\chi(s,b)],
\end{eqnarray}
with $t= -q^2$.
The second 
corresponds to
  the $U-$matrix approach \cite{umat}:
\begin{eqnarray}
  G(s,b) = U(s,b)/[1\ - \ i \ U(s,b)].
\end{eqnarray}
 where $U(s,b)= i \chi(s,b)=T_{born}(s,b)$.
 In \cite{trosh} it was shown that such a solution can  lead  to a nonstandard behaviour of the
  ratio
 $         \sigma_{el} / \sigma_{tot} \ \rightarrow \ 1, $
as $s\rightarrow \infty$.

These schemes  
lead to a decrease in the growth of  $\sigma_{tot}$,
but they predict very different energy regions for the onset of saturation.
  We show also \cite{cs1} that such a decrease heavily depends on the
  profile in impact parameter dependence of the Born amplitude.
In particular, if one takes the eikonal phase in the factorized form,
one  usually supposes that despite the fact that the energy dependence of
$h(s)$ can be a power of
$ h(s) \sim s^{\Delta}$,
the total cross section will satisfy the Froissart bound
$            \sigma_{tot} \leq \ a \ \log^2 (s)$.
We find in fact that
the energy dependence of the imaginary part of the amplitude
and hence of the total
cross section depends on the form of $f(b)$, {\it i.e.}, on
the $s$ and $t$ dependence of the slope of the elastic scattering
amplitude. For example, a power dependence on the impact parameter will
still violate the 
  Froissart-Martin bound after eikonalization.

\subsection{Saturation and non-linear equations }
 Different approaches of the saturation of  the scattering amplitude
try to connect it with the non-linear saturation processes which have been
  considered in a perturbative QCD context \cite{grib,mcler}.
Such processes lead to an infinite set of coupled
evolution equations in energy for the correlation functions of the
multiple Wilson lines.
In the approximation where the correlation functions for more than
two Wilson lines factorize, the problem reduces to the non-linear
Balitsky-Kovchegov (BK) equation \cite{BK}.

It is unclear how to extend these results to the non-perturbative region,
but one will probably obtain a similar equation. In fact we found simple
differential equations that reproduce either the $U$-matrix or the eikonal
representation.

  Let us  consider the saturation equation that has been used for many years
   in the various fields of physics - that is
 the logistic equation:
\begin{eqnarray}
  \partial N/\partial y = d \ N \ (1 - \lambda \ N) \label{cum-eq};
\end{eqnarray}
  This equation has a standard solution in the form
\begin{eqnarray}
  N = f(b) \  e^{ d\ \lambda \ y}/[1 \ + \ \lambda \ f(b) \ e^{ d \ \lambda \ y}] \label{cum}.
\end{eqnarray}
If, inspired by the BK results, we take $d=\alpha(0)-1$ with $\alpha(0)$  
the intercept of the Pomeron
and use the evolution variable
$y=\log s$,
with $\lambda=1$,   Eq. (\ref{cum}) coincides with the standard 
$U$-matrix representation
  of the scattering amplitude
obtained in the $U$-matrix formalism for $\Im U(s,b)=s^\Delta f(b)$.
  Hence, Eq. (\ref{cum}) extends the unitarized form
  of the scattering amplitude and is determined by the non-linear equation 
(\ref{cum-eq}).
 We can see that the asymptotic regime ($y=\log(s) \rightarrow \infty$)
  leads to a bound $1/\lambda$ for the amplitude, and gives the standard
BDL bound for   $\lambda=1$.
    We can use the same procedure \cite{cs7}
to obtain other unitarized forms of the
scattering amplitude.

  For example, from the non-linear equation
\begin{eqnarray}
     \partial N/\partial y \ = \ d \ (1\ - \lambda \ N) \ \log(1 \ - \ \lambda \ N),
\end{eqnarray}
we obtain a solution
\begin{eqnarray}
 N = \frac{1}{\lambda} \ [ 1 \ - \ e^{ - s^d / f(b)}] \label{cei}.
\end{eqnarray}
  The equation reproduces the extended eikonal (quasieikonal) form 
\cite{t-m}.
  With $\lambda=1$,  Eq. (\ref{cei}) coincides with the standard eikonal 
form of
  the scattering amplitude in the impact parameter representation.

We show in Fig. 1 and Fig. 2 the dependence of these non-linear equations 
on the size of the intercept of the Pomeron and on the  energy 
$S=\sqrt{s}$ in the case $\lambda=1$. 
  We see that the qualitative behavior of both equations is  very similar.
   The derivative in Eq. (\ref{cei}) has a larger maximum at low $s$ than
that in  Eq. (\ref{cum}). However, at large $s$, the situation in the opposite:
   the derivative in  Eq. (\ref{cum}) exceeds
   that of Eq. (\ref{cei}).

\begin{figure}[htb]
\includegraphics[width=.5\textwidth]{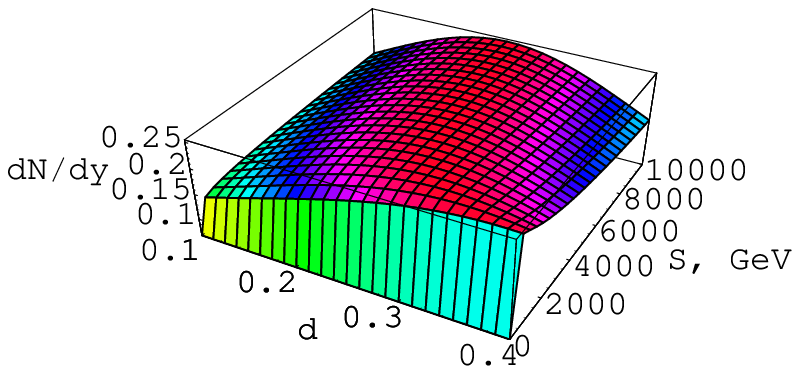}
\vspace{-5cm}
\begin{flushright}
\includegraphics[width=.5\textwidth]{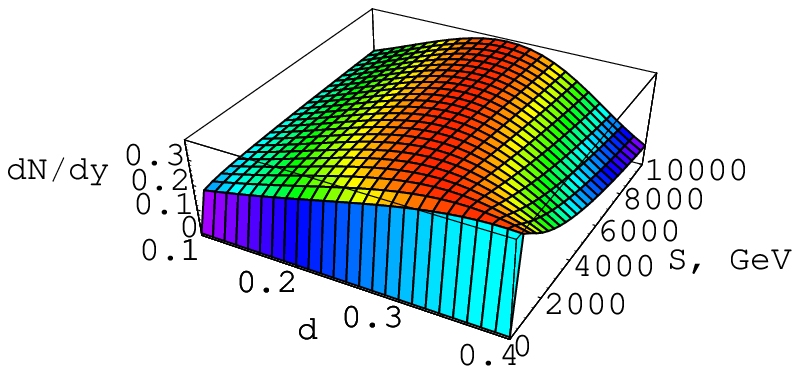}
\end{flushright}
\caption{ $dN/dy$ for Eq. (1.2) - left - and
  for Eq. (1.5) - right -.
}
\label{fig:largenenough}
\end{figure}

   We have shown that the most usual unitarization schemes could be recast into
  differential equations which are reminiscent of saturation equations
\cite{BK}.
  Such an approach can be used to build new unitarization schemes
  and may also shed light on the physical processes
underlying the saturation regime.

\begin{figure}[htb]
\begin{center}
\includegraphics[width=.8\textwidth]{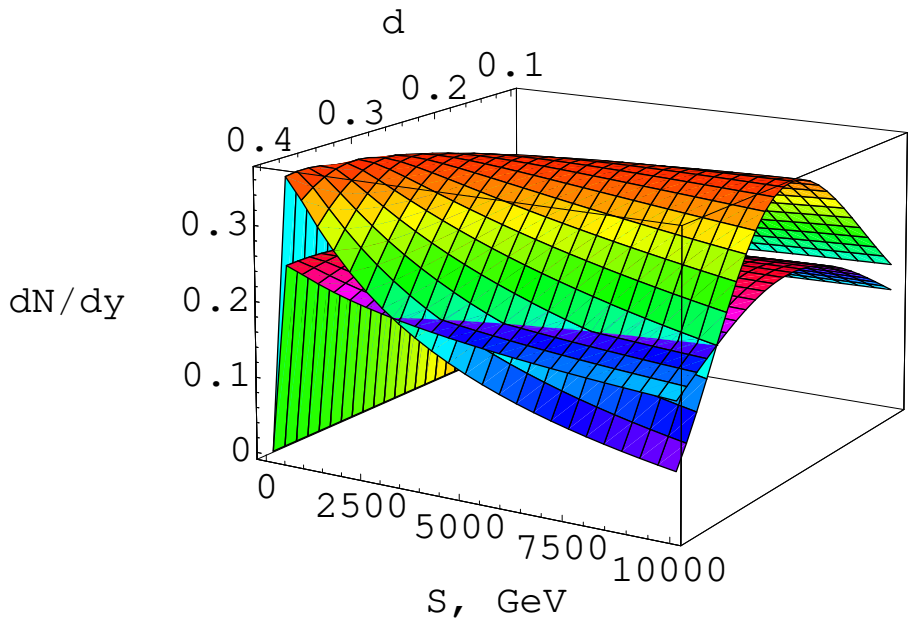}
\end{center}
\caption{$dN/dy$ for Eq. (1.2)  and
  for the Eq. ( 1.5)
}
\label{fig:largenenough2}
\end{figure}

 As an example of other unitarization schemes, the equation
\begin{eqnarray}
 \frac{\partial N}{\partial y} = dN_{bare}/dt \ [1-N^{2}] \label{nl-n2}
\end{eqnarray}
  has the unitarized solution
\begin{eqnarray}
 G(s,b)\ = \ \tanh[N_{bare}] . \label{tangh}
\end{eqnarray}

\subsection{Unitarization }

Let us now compare the behavior of the scattering amplitudes 
in different unitarization schemes. They are given by 
\begin{eqnarray}
  F(s,t) = i \int_{0}^{\infty} \ b\ db \ G(s,b) \ J_0 (q \ b),
\end{eqnarray}
  where $G(s,b) $ is the overlapping function which depends on the
  scheme and satisfies the BDL bound $G(s,b) \leq 1$.
 $ G(s,b)$ is a function of $ s, \ b, \ N_{bare}$,
and
\begin{eqnarray}
  N_{bare}(s,b) =i \int_{0}^{\infty} \ q\ dq \ F_{Born}(s,q) \ J_0 (q \ b).
\end{eqnarray}
   If the Born scattering amplitude is in the Regge form
\begin{eqnarray}
        F_{Born}(s,t) = i \ h \ s^d \ exp(R^2/2 \ t) \label{regge-t}
\end{eqnarray}
the corresponding form in the impact parameter representation is given by
\begin{eqnarray}
        F_{Born}(s,b) = i h \ \frac{s^d}{2 R^2} \ exp(-b^2/(4 R^2)).
\end{eqnarray}
    This Born amplitude will be the input of the different unitarization schemes.

  Following the Van Hove normalization \cite{VH}, 
the corresponding total cross section is
\begin{eqnarray}
 \sigma_{tot}(s) \ =     4 \pi  \ \int_{0}^{\infty} \ b \ [G(s,b)] \ db.
\label{tot0}
\end{eqnarray}

In Fig. 3, we show the total cross section for different unitarization schemes.
For all the schemes, the Born amplitude was taken in the simplest Regge form (\ref{regge-t}) 
 with the parameters corrsponding to the exchange of a hard Pomeron: 
$$ h \ = \ 0.1; \ \ \ R\ = \ 3 \ {\rm GeV}^{-1}; \ \  \ d \ = \ \alpha(0)-1 \ = \ 0.4$$
Here we want to compare the schemes directly. It is clear that 
if we make a fit to experimental data,
  the constant $h$ can be different for different unitarization schemes.

\begin{figure}[htb]
\begin{center}
\includegraphics[width=.8\textwidth]{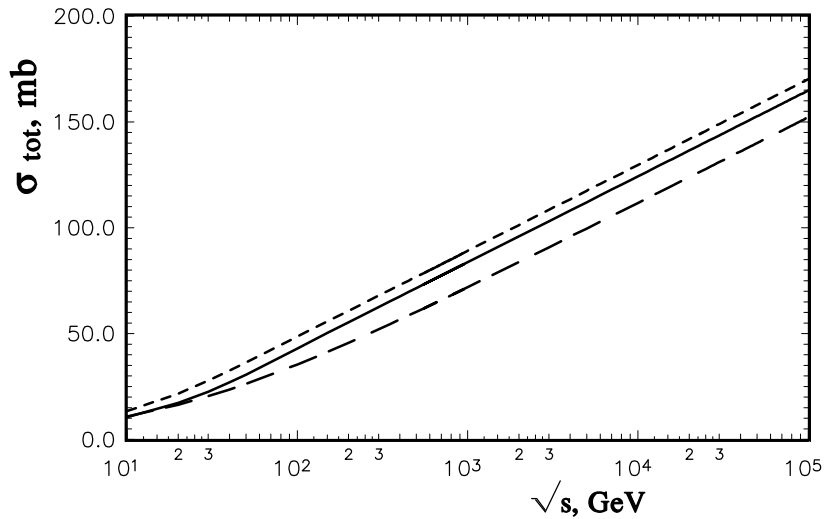}
\caption{Total cross sections for different unitarization schemes. 
The hard, long-dashed and short-dashed lines correspond to the eikonal, 
the $U$-matrix and the tanh forms
 of unitarization.
}
\end{center}

\label{fig:largenenough3}
\end{figure}
   We can see from Fig. 3 that at small energies ($\sqrt{s} \ = \  10 \ $GeV) the total cross sections
  are practically the same for all unitarization schemes. At this energy the unitarized amplitude practically
  reduces to the Born amplitude. When the energy grows, the schemes differ
most
   in an intermediate regime which corresponds to
  the energy region of approximately $30 \ $GeV $ \ \leq  \ \sqrt{s} \ \leq \ 200 \ $GeV. This regime
  corresponds to the growth of the overlapping function which is different 
in 
  different unitarization schemes. However, when we come to the asymptotic regime (with saturation of
  the BDL bound) the difference ceases and the growth is practically the same in a wide
   energy region.

 Hence, the asymptotics of all unitarization schemes are practically the same.
   Such a behavior  leads to two methods of analysing experimental data.
  One is connected with a different contribution of the secondary Reggeons 
 to the cross section. In this case, the asymptotics of the total cross sections
  can be exactly the same for different unitarization procedures.
 The  second solution is obtained  if we take the same contributions
   of the secondary Reggeons. The asymptotics of $\sigma_{tot} $ will then have 
   the same energy dependence but  some quantitative difference between the different
   unitarization schemes. In this case, the comparison of the predictions 
for the total cross sections
  at LHC energies from the different fitting procedures with different unitarization schemes
 will point out the true unitarization scheme and the true form
   of the non-linear equation.

\section{Conclusion}

 In the presence of a hard Pomeron \cite{clms2}, 
 the saturation effects
  can change the behavior of some features of
  the cross sections already at LHC
  energies.
    Hence, the problem of  unitarization of the scattering amplitude will be
     very important for the analysis of experimental data.

 Non-linear effects which work in the whole
  energy region  supply an acceptable growth of the total cross sections.
 Saturation leads to a relative growth
  of the contribution of  peripheral interactions.
 The most usual unitarization schemes
  can be recast into
  differential equations which are reminiscent of saturation equations.
  We can see that  the different forms of the non-linear equations
  lead to the different behavior of the scattering amplitudes at intermediate
 energies and to
  the same behavior at asymptotic energies.
  Further exploration of the unitarization schemes and of their connection with
  the future experimental data of LHC
  may  shed some light on the physical processes
underlying the saturation regime at the partonic level.

\section*{Acknowledgments} O.V.S. thanks the Chairmen of the Organized Committee
  of Conference C. Ktorides and R.~Fiore for the invitation and financial support.
O.V.S. acknowledges the support
of FRNS (Belgium) for visits
  to the University of Li\`ege where part of this work was done.
   We thank E. Martynov,
  S. Lengyel, G. Soyez and P.V. Landshoff for discussion.

\end{document}